\documentstyle[prl,aps,epsf,multicol]{revtex}
\begin{document}
\sloppy
\draft
\title{Phase transitions and noise crosscorrelations in a model of directed
polymers in a disordered medium}
\author{Abhik Basu \cite{bypur}}
\address{Centre for Condensed Matter Theory, 
Department of Physics, Indian Institute of Science,
Bangalore 560012, India\\and\\Poornaprajna Institute of Scientific Research,
Bangalore, India.}
\maketitle
\sloppy
\begin{abstract}
We show that effective interactions mediated by disorder between 
two directed polymers can be modelled
as the crosscorrelation of noises in the Kardar-Parisi-Zhang (KPZ)
 equations satisfied by the
respective free energies of these polymers.
When there are two polymers, disorder introduces attractive interactions
between them. We analyze the phase diagram in details and show that these 
interactions lead to new phases in the phase diagram. We show that, even 
in dimension $d=1$, the two directed polymers 
see the attraction only if the strength of the disorder potential exceeds 
a threshold value. We extend our calculations to show that if there are 
$m$ polymers in the system then $m$-body interactions are generated in the
disorder averaged effective free energy.
\end{abstract}


\section{introduction}
\subsection{Background}
Studies of phase transitions in presence of disorder have opened up new 
problems in statistical mechanics. Phase transitions in spin glasses and 
polymers in disordered media are typical examples  
\cite{stanley,halpin,mezard,barat,natter}. Various attempts have been 
made to understand
such transitions see, e.g., Ref.\cite{parisi,suta}. In Ref.\cite{parisi}
replica trick has been used while in Ref.\cite{natter,suta,fns} dynamic 
renormalisation group has been used and the relevant exponents (see Section II)
have been calculated. The nature of overlaps, if there are more than one polymer,
in the system has also been examined \cite{suta}. However whether directed 
polymers (DP)
can interact via disorder in the absence of any {\em direct} interaction has not yet
been studied. In this paper we ask: If there are more than one DPs in the
disordered medium, can they interact through disorder in absence of any
direct mutual interaction? We find that 
in the absence of any direct interactions disorder mediates effective 
attractive interactions between the DPs which lead to binding transitions 
between them, which are  qualitatively similar to \cite{suta}.

A directed polymer in $d+1$ dimensions is just a directed string stretched 
along one
particular direction with free fluctuations in all other $d$ transverse 
directions. The Hamiltonian of a directed polymer in a quenched random
potential is
\begin{equation}
{H\over k_BT}=\int_0^t d\tau [{\nu\over 2}({d{\bf x}\over\ d\tau})^2 + 
{\lambda\over\ 2\nu} V[{\bf x}(\tau),\tau]],
\label{hamil}
\end{equation}
where $x(t)$ is the $d$-dimensional transverse spatial coordinate
of the directed polymer at length $t$. The first term is the energy due
to transverse fluctuations (elastic energy) and the second one is the
potential energy due to disorder. In a random environment there is a 
competetion between the potential energy due to randomness and the elastic energy.
The random potential $V$ is chosen to be Gaussian-distributed,
zero mean with a correlation given by
\begin{equation}
\langle V({\bf k},t)V({\bf k'},t')\rangle =2D\delta^d ({\bf k+k'})\delta (t-t').
\end{equation} 
The elastic term $\nu({d{\bf x}\over d\tau})^2$ attempts to smoothen out the DP
(minimum $\nu({d{\bf x}\over d\tau})^2$ everywhere), while the disorder potential
favours the DP to take a rough profile so that the DP can go through the low
energy paths. At a low enough temperature the second option is energitically
favourable and a disorder-dominated super diffusive phase is produced
\cite{halpin,natter,parisi,kpz,mezard1}.
 At $d>2$ a transition is observed from low temperature
strong disorder phase to high temperature smooth phase 
\cite{natter,derrida,doty}
, which is described by an unstable fixed point to $O(\epsilon)$
in a dynamic RG calculation \cite{natter,kpz}.
There is a one-to-one mapping between the problem of a directed polymer
in a random medium and the nonequilibrium surface growth problem described 
by the Burgers/Kardar-Parisi-Zhang (KPZ) equation.
Burgers equation is the simplest non-linear generalisation of the diffusion
equation \cite{burg}. This equation is used to describe diverse phenomena:
structure formation in astrophysical situations, turbulence etc. This 
can be mapped into the Kardar-Parisi-Zhang (KPZ) equation which is a prototype 
model for nonequilibrium growth surfaces:
\begin{equation}
{\partial h\over\partial t}+{\lambda\over\ 2}(\nabla h)^2 =\nu\nabla ^2 h +\eta.
\label{kpz}
\end{equation}
The Gaussian noise satisfies
\begin{equation}
\langle \eta({\bf x},t)\eta ({\bf x}',t')\rangle=D\delta ^d({\bf x-x}')
\delta (t-t').
\end{equation}
The long wavelength and long time properties of this 
equation have been studied extensively using dynamic renormalisation
group method \cite{stanley}.

Interestingly,  this equation can be transformed into a linear equation
by Cole-Hopf transformation
\cite{stanley}:
\begin{equation}
h(x,t)=(2\nu/\lambda) \ln\, W\;\equiv k_BT \ln \,W.
\end{equation}
The resultant linear equation is a diffusion equation with a multiplicative
noise:
\begin{equation}
{\partial W\over\partial t}=\nu \nabla^2 W +{\eta\over\ 2\nu} W
\end{equation}
The partition function $Z$ of a directed polymer satisfies the above 
equation with $t$ being the coordinate parametrising the length of the 
polymer. It immediately follows that the free energy of the DP, 
$h\equiv k_BT\ln\,Z$ satisfies the KPZ Eq. \ref{kpz}.
The corresponding Hamiltonian is the Hamiltonian of a directed polymer as 
given in \ref{hamil}.
The dictionary between 
the surface growth problem described by the KPZ equation and the
directed polymer problem described the Hamiltonian (\ref{hamil}) is
as follows: The temperature scale of the polymer has been set to one;
hence the elastic modulus of the polymer is given by 
$c\equiv {1\over\ 2\nu},$
and the height variable $h$ of the KPZ equation gives the free energy
of the polymer problem. The probability distribution of the quench
random potential $V({\bf x},t)$ is same as that of the noise $f({\bf x},t)$
in the KPZ equation \ref{kpz}. The relevant
exponents are $\chi$ and $\zeta=1/z$. $\chi$ describes the free energy
fluctuations $f\sim t^{\chi/z}$ ($\chi$ is also the roughness exponent of the
height field $h$. $z$ is given by  
$\langle [x(t)-x(0)]^2\rangle \sim t^{2/z}$ and is
also the dynamic exponent of $h$. Due to the Galilean invariance of the 
KPZ equation there is an exact exponent relation
\begin{equation}
\chi+z=2.
\end{equation}
\subsection{Results}
In this paper we investigate if there are two DPs in a random medium,
whether  the medium can induce interactions leading to
phase transitions, in absence of any direct mutual interactions
between the polymers. We find that
disorder indeed induces effective attractive 
interactions between the a polymers which
causes  phase transitions in the system. In terms of the KPZ descriptions
this implies that noises in the
individual KPZ equations satisfied by the free energies of the respective
polymers have nonzero crosscorrelations. In section II, we show that
if there are two polymers in the system, then crosscorrelations of the random
potentials seen by them may lead to a binding-unbinding transition of
them. We analyze the phase diagram and show that new phases appear, which do
not exist when there is only one polymer in the medium. We calculate the
crossover exponent (defined below). In Section III, we
generalise our calculations and show that effective $m$-polymer
interactions are also generated. We summarise our results in Section IV.

\section{Two directed polymers in a random medium}
\subsection{Model}
In absence of any direct interactions the Hamiltonian of two DPs in a random
medium becomes just of the sum of the individual Hamiltonians $H_1$ and $H_2$:
\begin{equation}
H=H_1+H_2=\int_0^{\tau} [{\nu \over 2}({d{\bf x}_1\over dt})^2+
 {\nu \over 2}({d{\bf x}_2\over dt})^2+V_1({\bf x}_1,t)+V_2({\bf x}_2,t)],
\label{ham2}
\end{equation}    
In two papers \cite{suta} Mukherjee and Bhattacharjee
 have studied overlap of directed polymers in a random medium
(see also \cite{derrida1} for a related problem on a lattice).
They calculated the overlap of polymers by introducing new interactions 
in the Hamiltonian (\ref{ham2}) and calculated the scaling behaviour using a
dynamic renormalisation group approach. Their modified Hamiltonian is given by
\cite{suta}
\begin{equation}
H_m=\sum_{i=1}^m H_i+(\lambda/ 2\gamma)v_m\int_o^td\tau\Pi_{i=1}^{m-1}
\delta[{\bf x}_{i,i+1}(\tau)],
\label{newham}
\end{equation}
where $H_i$ is the Hamiltonian (\ref{hamil}) for the $i$th polymer and
${\bf x}_{i,i+1}={\bf x}_i-{\bf x}_{i+1}$. 
The presence of the additional $\delta$-functions interactions ensure that
the overlap between $N$-polymers 
is nonzero. Here $H_m$ is the Hamiltonian (\ref{newham}). 
However, due to the additional terms
in their Hamiltonian representing the overlap, the equation that is  
satisfied by the corresponding total free energy is not quite the usual 
KPZ equation; it has an additional term as noise:
\begin{equation}
{\partial h\over\partial t}=\sum_{j=1}^m [\gamma \nabla^2 _j h +{\lambda\over 2}
(\nabla _j h)^2 ]+g_o,
\label{newkpz}
\end{equation}
where
\begin{equation}
g_o=\sum_{j=1} ^m [\gamma V(x_j,t)+v_m \Pi _{j=1}^{m-1}
\delta [x_{j,j+1}(\tau)].
\end{equation}
Due to this unusual looking noise term, although Eq.(\ref{newkpz})
looks like a higher ($md$) dimensional KPZ equation, it is not really so.
With this modified Hamiltonian/KPZ equation the 
crossover exponents $\phi_m$ for overlap of $m$ chains for $v_m\rightarrow
0$ have been calculated in Ref.\cite{suta}. We however do not include
additional interactions and work with the Hamiltonian \ref{ham2}.

\subsection{Effective interactions between two directed polymers: The phase
diagram}
Let us again consider the Hamiltonian for two DPs in a disordered medium:
\begin{equation}
H=H_1+H_2=\sum_{i=1,2}\int_o^t d\tau [{\nu\over 2}({d{\bf x}_i\over d\tau})^2
+{\lambda\over 2\nu}V[{\bf x}_i(\tau),\tau]],
\end{equation}
where $H_1$ and $H_2$ are the Hamiltonians for the two DPs respectively,
${\bf x}_1$ and ${\bf x}_2$ are their transverse fluctuations.
We take the total Hamiltonian to be additive (i.e. a sum of $H_1$ and $H_2$)
as there are no direct mutual interactions between the polymers. $V_1=V({\bf x}_1)$ 
and $V_2=V({\bf x}_2)$ are the potential energies.  The total
partition function $Z$ is the product of the individual partition functions:
$Z=Z_1\,Z_2=\int {\cal D}x_1 {\cal D}x_2\, \exp[-\beta(H_1+H_2)]$. 
$Z_1$ and $Z_2$ satisfy
\begin{eqnarray}
{\partial Z_1\over\partial t}&=&\nu\nabla^2 Z_1+Z_1V_1, \\
{\partial Z_2\over\partial t}&=&\nu\nabla^2 Z_2+Z_2V_2,
\end{eqnarray} 
while the free energies $h_1$ and $h_2$ satisfy (as $Z_{1,2}=\exp (-\beta h_{1,2})
$)
\begin{eqnarray}
{\partial h_1\over\partial t}&=&{\lambda\over 2}(\nabla h_1)^2+\nabla^2 h_1+
f_1, \\
{\partial h_2\over\partial t}&=&{\lambda\over 2}(\nabla h_2)^2+\nabla^2 h_2+
f_2. 
\end{eqnarray}

Let us calculate 
(the (cross)-correlation function of the transverse fluctuations
of the two polymers)
$\overline {{\int{\cal D}{\bf x}_1{\cal D}{\bf x}_2\, {\bf x}_1(t_1){\bf x}_2(t_2)(
\exp [-\beta (H_1+H_2)]}/Z}$. A `$\overline{\hskip0.5cm}$' indicates 
averaging over disorder realisations. 
We may evaluate it in lowest order in $\lambda$ in a perturbation
expansion. Expanding the Boltzmann factor in powers of $\lambda$, we obtain
\begin{eqnarray}
\overline {{\int{\cal{D}}{\bf x}_1{\cal D}{\bf x}_2\,{\bf x}_1 {\bf x}_2\exp [-\beta (H_1+H_2)]}/Z}&=& 
\int\Pi {\cal D}{\bf x}_i\,{\bf x}_i
\exp[-\beta (\sum_{i=1,2}{d{\bf x}_i^2\over d\tau}) \nonumber \\
&&[1+\lambda {\overline V_1}_c+\lambda^2 {\overline V_2}_c
+\lambda^2 {\overline {(V_1^2/2)}}_c+\lambda^2 {\overline {(V_2^2/2})}_c+
\lambda^2 {\overline {(V_1V_2)}}_c+....],
\end{eqnarray}
 In the right hand side a subscript $c$ indicates cumulants, i.e., contributions only
from the connected diagrams should be considered. We take 
\begin{eqnarray}
{\langle {V({\bf x}_{1},t)V({\bf x'}_{1},t')}}&=&
2D_1\delta ({\bf x}_{1}- {\bf x'}_{1})\delta(t-t'), \nonumber \\
{\langle {V({\bf x}_{2},t)V({\bf x'}_{2},t')}}&=&
2D_2\delta ({\bf x}_{2}- {\bf x'}_{2})\delta(t-t'). \nonumber \\
\label{autocor}
\end{eqnarray}
These we call as `autocorrelations' in a sense 
that in Eq.\ref{autocor} coordinates refer to the same DP. 
It is clear that unless $\langle V_1({\bf x_1},t)V_2({\bf x_2},t)\rangle $
is nonzero the crosscorrelation function (as defined above)
vanishes identically. We choose
\begin{equation}
\langle V_1({\bf x}_1,t)V_2({\bf x}_2,t')\rangle=2\tilde {D}
\delta({\bf x}_1-{\bf x}_2)\delta(t-t'),
\label{crosscor}
\end{equation}
where $\tilde{D}$ may differ from $D_1$ or $D_2$ as the two polymers may be 
distinguishable. We want 
to investigate the effect of this on the phase diagram
of the polymers. We give a physical interpretation of this shortly below.
Let us see what this condition means in terms of the KPZ description: The free 
energies $h_1$ and $h_2$ of the two polymers satisfy usual KPZ equations.
Without any loss of generality we put $D_1=D_2=D$. We have (see Eq.\ref{autocor})
\begin{eqnarray}
\langle f_1 (k,t) f_1(k',t')\rangle =2D \delta ^d (k+k')\delta (t-t'), \\
\langle f_2 (k,t) f_2(k',t')\rangle =2D \delta ^d (k+k')\delta (t-t'). 
\end{eqnarray}
Such a choice, with $\langle f_1 f_2\rangle =0$  would automatically 
guarantee that $h_1$ and $h_2$ represent the free energies of two
{\em identical, mutually noninteracting} polymer in a disordered medium. 
It is clear that having a nonzero $\langle V_1 (k,t) V_2(k',t')\rangle$
implies a nonzero $\langle f_1 (k,t) f_2(k',t')\rangle$:
\begin{equation}
\langle f_1(k,t) f_2(k',t') \rangle =2\tilde{D} \delta^d(k+k') \delta (t-t')
\end{equation}
If $\langle f_1(x,t)f_2(x',t')\rangle = 0$ or
 $\langle V_1(x,t)V_2(x',t')\rangle = 0$ then after averaging over
disorder the effective Hamiltonian is just the sum of two noninteracting 
single chain Hamiltonians. Obviously there is no attractive interaction
between the two polymers.
However since both the polymers are in the same random medium, there are
of course correlations between the random potentials seen by the two
polymers, which, as we will see, mediate interactions between the
polymers. Equivalently saying,
the polymers interact through the disordered medium.
Mathematically, on averaging over the disorder distribution, 
due to the non-zero cross-correlations of the noises, a new term $\propto
\tilde{D}\delta ^d (x_1(\tau)-x_2(\tau))$ is generated in the 
effective Hamiltonian.  This new `effective potential' can be
interpreted as an attractive interaction felt by one of the polymers when it 
is in contact with the other (see Fig.1). This will cause, as we have seen before,
quantities like $\langle {\bf x_1x_2}\rangle$ to be nonzero for certain
strength of the interactions.  We show below that this effective attractive
interaction leads to a phase transition involving the two DPs.
\begin{figure}[htb]
\epsfxsize=4cm
\epsfysize=5cm
\centerline{\epsffile{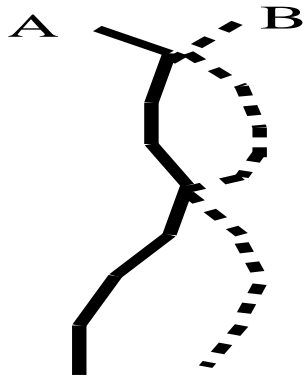}}
\caption{A schematic diagram showing an effective contact interactions 
between two polymers
that arises due to  crosscorrelation of the random potential (see text).}
\end {figure}

The total free energy $F$ of the two DPs after averaging over disorder
distribution is a function of the couplings $\tilde{D}$:$F\equiv 
F(\tilde{D},t)$. We define, as in \cite{suta} the order parameter as the 
derivative  of the quenched free energy with respect to the appropriate coupling
constant:
\begin{equation}
q(t)={1\over\ t} \int_0 ^t d\tau \delta (x_1 (\tau)-x_2 (\tau))
=-{1\over\ t} {dF (\tilde{D},t)\over\ d\tilde{D}}|_{\tilde{D}=0},
\end{equation}
Here $F (\tilde{D}, t)$ is the scaling part of the effective free energy for 
the for the two polymers. Following Ref.\cite{suta} we start with a scaling form
\begin{equation}
F(\tilde{D}, t)=t^{\chi/z} f(\tilde{D} t^{-\phi_2/z}).
\end{equation}
Here $\phi_2$ is the crossover exponent. This gives \cite{suta}
\begin{equation}
q=t^{\Sigma_2} Q(D t^{-\phi_2/z})
\end{equation}
where $\Sigma_2 = (\chi-\phi_2-z)/z$. 
We calculate the crossover exponent
in a one-loop dynamic renormalisation group calculation.

\begin{figure}[htb]
\epsfysize=7cm
\centerline{\epsffile{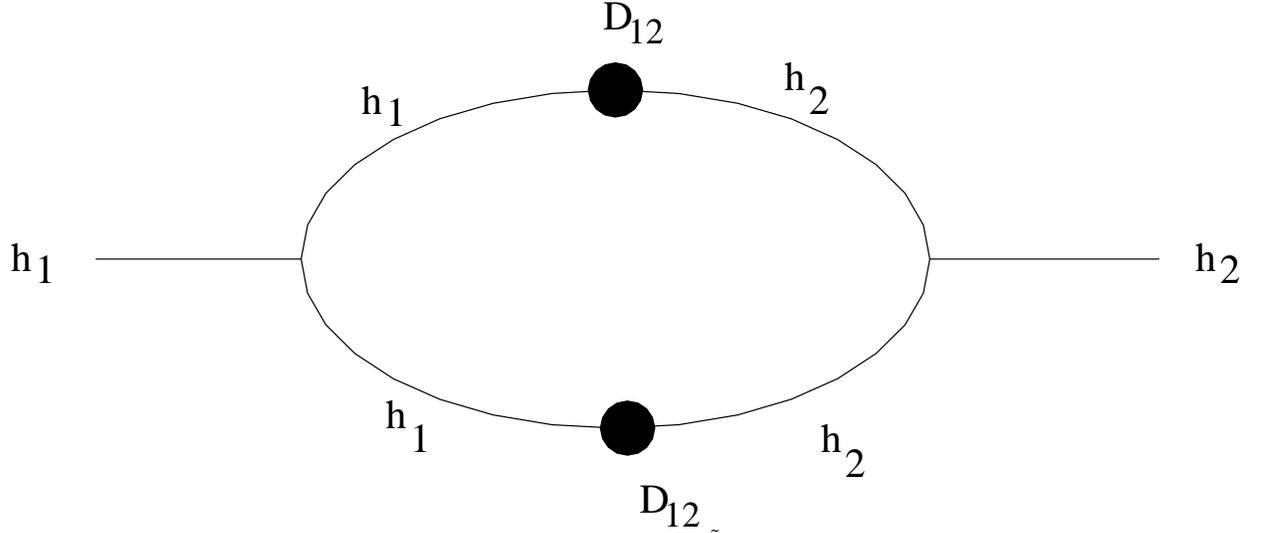}}
\caption{The 1-loop diagram, contributing to the renormalisation of $\tilde{D}$;
 continuous lines indicate  $h_1$ or $h_2$ lines as mentioned in the figure
and the small filled circles indicate bare $\tilde{D}=D_{12}$.}
\label{fig1}
\end{figure}

\begin{figure}[htb]
\epsfxsize=12cm
\centerline{\epsffile{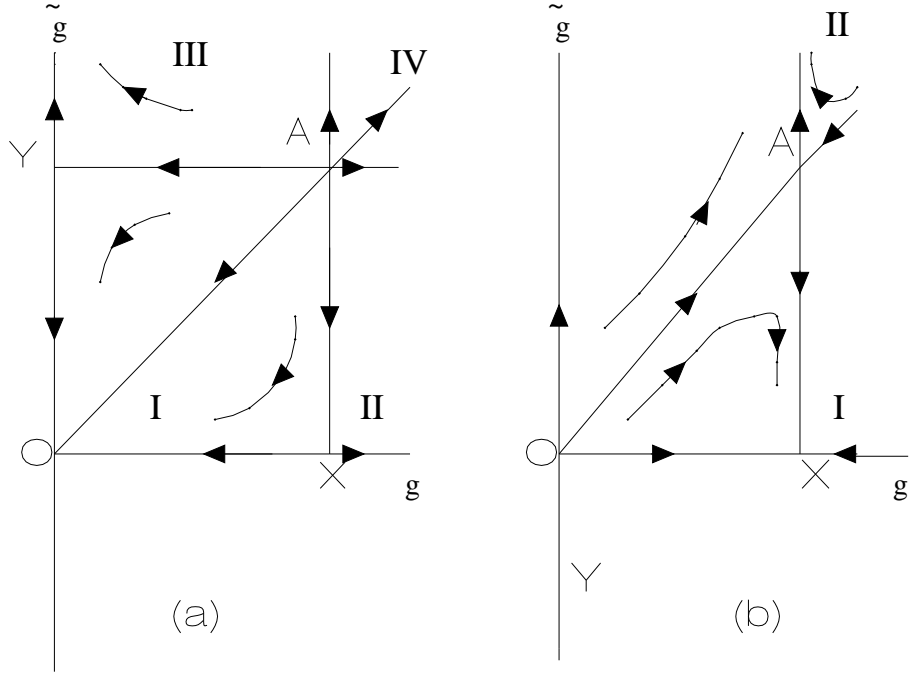}}
\caption{A diagram indicating the flow around the fixed points in $g-\tilde{g}$
plane:(a)$d=2+\epsilon$,(b)$d=1$.}
\label{fix}
\end{figure}

Renormalisation group flow equations for the parameters of a single KPZ equation
(namely, for $\nu$, $D$, and $\lambda$) have already been 
calculated \cite{stanley}. We present
the calculation for $\tilde{D}$: The following diagram will contribute at the
one-loop level (Fig.\ref{fig1}).
The one-loop integral, after frequency integration becomes $\tilde{D}^2 \int 
{d^dq \over\ (2\pi)^d} {1\over\nu^3 q^2}\sim \Lambda^{d-2}$, where $\Lambda$
is some momentum scale coming from the lower limit of the one-loop integral. 
We see that this integral has the same infra-red
behaviour as the usual one-loop integral that comes in for the renormalisation
of the noise correlations in a single KPZ equation. Under 
rescaling of space and time, different parameters scale according to their
naive dimensions:$\nu\rightarrow b^{z-2}\nu, \lambda\rightarrow b^{z+\chi-2}
\lambda, D\rightarrow b^{z-d-2\chi}D$ and $\tilde{D}\rightarrow 
b^{z-d-2\chi}\tilde{D}$. The flow equation
for different coupling constants are:
\begin{eqnarray}
{d\nu\over\ dl} &=& [z-2 +k_d g(2-d)/4d]\nu, \\
{d\lambda\over\ dl} &=& [z+\chi -2]\lambda, \\
{dD\over\ dl}&=&[z-d-2\chi+gk_d/4]D, \\
{d\tilde{D}\over\ dl}&=&[z-d-2\chi+\tilde{g}k_d/4]\tilde{D},
\end{eqnarray}
where $a$ is the smalest length scale in the problem, and $k_d^{-1}=2^{d-1}
\pi^{d/2}\Gamma(d/2)$ and $g=(a/\pi)^{2-d}D\lambda^2/\nu^3$ and $\tilde{g}=
(a/\pi)^{2-d}\tilde{D}\lambda^2/\nu^3$ are the dimensionless coupling constants.
It is easy to obtain the flow equations for the coupling constants:
\begin{eqnarray}
{dg\over\ dl}&=&(2-d)g+k_d{2d-3\over\ 2d}g^2, \\
{d\tilde{g}\over\ dl}&=&\tilde{g}(2-d+k_d{\tilde{g}\over 4}-3k_dg{2-d\over\ 4d}).
\end{eqnarray}
These equations have fixed point solutions $O:[g_0=0\,\tilde{g}=0], 
X:[g=2d(d-2)/k_d (2d-3)\equiv g_c,\,\tilde{g}=0]$,
$Y:[g=0,\,\tilde{g}=d-2],\, A:[g=g_c,\,
\tilde{g}=g_c]$.  We show the flows in a fixed point diagram in Fig.\ref{fix}.

 Among these values, $O:(0,0)$
 for $d\equiv 2+\epsilon$ are stable and correspond
to Gaussian polymers. At $d=1$, $g=g_c$ and $\tilde{g}=\tilde{g}_c$ are the
stable fixed points implying that any small amount of disorder makes the polymer
non-Gaussian.
At $d=2+\epsilon,\,Y\equiv (g=0,\tilde{g}=d-2),\,X\equiv (g=g_c,\tilde{g}=0),\,
A\equiv (g=g_c,\tilde{g}=g_c)$
are all unstable. A is unstable along both the directions, X is unstable along $g$ 
direction, and Y is unstable along $\tilde{g}$ direction. 
Hence they indicate second order phase transitions. At $d=1$, A is still unstable
along $\tilde{g}$ direction. Y takes a negative ordinate, reflecting presumably
a bound state not describable by a fixed point. So if one moves along AX one 
always encounters a second order phase transition. The disorder averaged
Hamiltonian has potential terms of the forms $-g\delta({\bf x}_1-{\bf x}_1'),
-g\delta({\bf x}_2-{\bf x}_2'), -\tilde{g}\delta({\bf x}_1-{\bf x}_2)$. The
last term comes from the crosscorrelation effects. Due to the negative sign 
it is attractive in nature. Recall that in $d=1$ X is stable along $g$ 
direction indicating that any small amount of disorder is relevant in
$d=1$. However, X is unstable along $\tilde{g}$ direction which shows that
unless the disorder strength exceeds a minimum value, the two
polymers don't attract each other. We obtain the following phases in the phase
diagram (see Fig.\ref{fix}) characterised by stable fixed points:
\begin{enumerate}
\item In Fig.\ref{fix}(a), i.e., for $d=2+\epsilon$
\begin{enumerate}
\item free Gaussian polymers characterised by the stable fixed point O(0,0)
(denoted by I).
\item free `strong disorder/KPZ' polymers, i.e., polymers don't attract each
other but their individual fluctuations are influenced by the disorder
(denoted by II).
This phase is characterised by a stable fixed point of the form $(G_1,0)$,
not accessible in perturbation theory.
\item bound Gaussian polymers: individual polymer fluctuations are unaffected
by disorder but they are bound due to the effective attarctive interaction 
(denoted by III).
This phase is characterised by a stable fixed point of the form $(0,\tilde{G}_1
)$, not accessible in perturbation theory.
\item bound KPZ polymers: individual polymer fluctuations are affected
by disorder, also they form a bound pair due an effective attractive 
interaction (denoted by IV). This phase is characterised by a stable 
fixed point of the $(G,\tilde{G})$ again not accessible in perturbation theory.
\end{enumerate}
\item In Fig.\ref{fix}(b), i.e., for $d=1$,
\begin{enumerate}
\item free `strong-dosorder/KPZ' polymers described by stable fixed point
(2,0) (denoted by I), Exponents are known {\em exactly}.
\item bound KPZ polymers (denoted by II) by a stable fixed point not accessible
in perturbation theory.
\end{enumerate}
\end{enumerate}
Note that in Fig.\ref{fix}, the line $g=\tilde{g}$ is invariant under the RG 
transformation; this is just a reflection of the fact that if the two DPs are
indistinguishable, they remain so in a coarse grained description.

In \cite{suta}
overlaps have been calculated at $(g=g_c,\tilde{g}=g_c)$. We, however, investigate
the nature of the phase transitions  considering $\tilde{g}$ 
as the ordering field at $g=\epsilon$ and $g=0$. Hence our definitions of order 
parameter is given by
\begin{equation}
q\sim {dF\over dD^*}|_{D^*=0}.
\end{equation}
where $D^*$ is an effective coupling constant which are functions of $\tilde{g}$.
In other words 
we wish to calculate the crossover exponent ${\phi}_2$ for the two
polymers at the fixed point $(g_c,0)$ in the $g-\tilde{g}$ plane. It is given by
\begin{equation}
\phi_2=2\chi+d-z+\delta_{12}=0
\end{equation}
Hence we obtain $\Sigma_2=(\chi-z)/z$ in all dimension. In particular
in $d=1$, $\Sigma_2=-1/2 $ and in $d=2+\epsilon$, $\Sigma_2=-1$ (since
$\chi=0$ in $O(\epsilon)$). The length scale exponent $\nu$ at the unstable
fixed points A and Y in Fig.\ref{fix}(a) (i.e., for $d>2$) is same,
however at O in Fig.\ref{fix}(b) (i.e., $d=1$) it is different from A.

\section{Effective interactions involving arbitrary number of polymers}
In this section, we generalise our previous calculations to show that 
when there are $m$ DPs the 
effective free energy can have any arbitrary $n$-polymer ($n\leq m$)
interactions. We show that these
interaction terms naturally arise in the free energy after disorder averaging.
Let us examine the quantity
 $\int\Pi_{i=1,m} {\cal{D}}{\bf x}\, {\bf x}_i \exp [-\beta H]$
where $H=H_1+...+H_{m}$ sum of the Hamiltonians of $m$ polymers
[each being same as (\ref{hamil})]. Following our calculations of the previous
Section we write
\begin{equation}
\overline{\int\Pi_i {\cal {D}}{\bf x} {\bf x}_i \exp(-\beta H)/Z}
=\int\Pi {\cal D}{\bf x} {\bf x}_i 
\exp(\sum_i {d{\bf x}_i^2\over d\tau})[1+...+\lambda^{m} (V_1...V_{m})]
\end{equation}
It is clear that $m$-point crosscorrelation function of transverse fluctuations
is nonzero is nonzero only if $\langle V_1...V_{m}\rangle_c$
is nonzero. We calculate $\langle V_1...V_m\rangle$ 
to $O(\lambda^m)$ ($\equiv$ one-loop)
in the KPZ description of the problem. As we shall see later,
\begin{equation}
\langle V_1({\bf k_1},t_1)...V_{m}({\bf k_{m}},t_{m})\rangle
\propto D_m \delta ({\bf k_1}+...+{\bf k_{m}})\delta(t_1-t_{m})...
\delta(t_{m-1}-t_m).
\end{equation}
We find out the scaling dimension of $D_{m}$ in a one-loop approximation.
We examine below the nature of the transition due to $D_m$ an effective
interaction which makes many point correlation functions $\langle x_1...x_m
\rangle$ nonvanishing. Now the effective free energy (averaged over random 
potential) of the 
$m$-polymers  is a function of all these effective couplings:
\begin{equation}
F\equiv F(\tilde{D},...,D_{m}).
\end{equation}
Following the usual definiton of order parameter as the derivative of the 
free energy with respect to the appropriate coupling constant we obtain 
\cite{suta}
\begin{equation}
q_{m}=-{1\over t}{dF(D_{m} t^{-\phi_{m}/z})\over d{D}_{m}}|
_{{D}_{m}=0}.
\end{equation}

It is easy to see that, in our approach,
each of the $m$ polymers' free energy will satisfy the usual KPZ equation
separately.
As in Section II we assume  noises present in these $m$ KPZ equations will
have non-zero cross-correlations between them:
\begin{eqnarray}
{\partial\over\partial t}h_1+{\lambda\over\ 2}(\nabla h_1)^2&=&\nu\nabla^2 h_1+f_1\\
{\partial\over\partial t}h_2+{\lambda\over\ 2}(\nabla h_2)^2&=&\nu\nabla^2 h_2+f_2\\
....................&&............................\nonumber \\
{\partial\over\partial t}h_{m}+{\lambda\over\ 2}(\nabla h_{m})^2&=&
\nu\nabla^2 h_{m}+f_{m},\\
\end{eqnarray}
with noise correlations given as in Section II
\begin{eqnarray}
<f_i(k,t)f_i(k',t')>&=&D\delta(k+k')\delta(\omega+\omega ') \\
<f_i(k,t)f_j(k',t')>&=&\tilde{D}\delta(k+k')\delta(\omega+\omega ');i\neq j\\
\end{eqnarray}

We define the relevant order parameter as
\begin{equation}
q_{m}=-{1\over\ t}\int_0 ^t d\tau\langle\Pi_{i=1}^{m-1}\delta(x_i(\tau)-x_{m}
(\tau))\rangle.
\end{equation}
We note that even though our definition of order parameter is actually
different from the defined in \cite{suta}, physically it measures the same
thing. For $m=2$ the two definitions are identical. It is evident that
$\langle V_1...V_{m} \rangle\propto\langle f_1...f_{m}\rangle $.

\subsection{Effective interactions between three polymers}
Let us consider a situation where we have three polymers.
Due to the Gaussian statistics
of the random potential, $\langle h_1({\bf x_1},t)h_2({\bf x_2},t)
h_3({\bf x_3},t)\rangle$ is zero at the bare level. However, due to the
nonlinearity in the equation,
$\langle h_1({\bf x_1},t)h_2({\bf x_2},t)h_3({\bf x_3},t)\rangle$ is
nonzero at the one-loop level. In Fig.2 we show the one-loop diagrams. 
\begin{figure}
\epsfysize=6cm
\centerline{
\epsffile{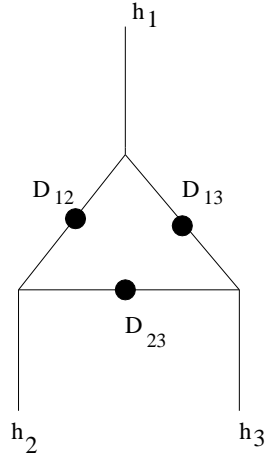}
}
\caption{A diagram showing all the 1-loop contributions to $\langle h_1 h_2 h_3
\rangle$. The filled circles refer to $D_{12}=D_{13}=D_{23}=\tilde{D}$.}
\end{figure}
Each of the above one-loop diagram scales as 
\begin{equation}
\sim\tilde{D}^{3} \lambda^3 \int {d^dq\over\ \nu^5 q^4}\sim
 \tilde{D}^{3} \lambda^3 \Lambda^{d-4}\sim {g_c}^{5/3}\Lambda^{d-4},
\label{anodi3}
\end{equation}
where $\Lambda$ is the lower cutoff of the momentum integral. We see that even 
though the bare noise is Gaussian the effective noise has nonzero third
cumulant (which arises due to the nonlinear term in the KPZ equation):
\begin{equation}
\langle f_1({\bf k}_1,t_1)f_2({\bf k_2},t_2)f_3({\bf k_3},t_3)\rangle
=D_3 \delta({\bf k_1}+{\bf k_2}+{\bf k_3})\delta(t_1-t_2)\delta(t_1-t_3)
\end{equation}
Naive dimension of $D_3$ is $z-2d-3\chi$ and anomalous dimension is $4-d$
(see Eq.\ref{anodi3}). Hence
\begin{equation}
\phi_3=-(z-2d-3\chi+4-d)=-(4-3d)=3\epsilon.
\end{equation}
at $d=2+\epsilon$ and (since $z=2+O(\epsilon^2)$ and $\chi=O(\epsilon^2)$)
consequently
\begin{equation}
\Sigma_3=(\chi-\phi_3-z)/z=(-3\epsilon-2)/2
\end{equation}
at $d=2+\epsilon$. We notice that $\Sigma_3<0$ indicating that 
in the thermodynamic limit i.e., when $t\rightarrow \;\;\infty$,
$q_3\sim t^{\Sigma_3}\sim \mid T-T_c\mid ^{-\nu\Sigma}\rightarrow 0$ as
$T\rightarrow T_{c^-}$. Here $\nu$ is the correlation length exponent.

\subsection{Effective interactions between four polymers}
Let us consider a situation when we have four polymers in the medium. We
define the order parameter
\begin{equation}
q_4\equiv -{1\over t}\int_o^td\tau \langle \delta ({\bf x}_1-{\bf x}_2)
\delta({\bf x}_1-{\bf x}_3)\delta({\bf x}_1-{\bf x}_4)\rangle
\end{equation}
Similar to our analysis of three polymers, we calculate 
$\langle f_1({\bf x}_1,t_1)f_2({\bf x_2},t_2)f_3({\bf x_3},t_3)f_4({\bf x_4},
t_4)\rangle$ in a one-loop perturbation theory. 
The one-loop diagrams are shown in Fig.3. The one-loop integrals scale as
$\sim {\tilde{D}^4\over\nu^7} \Lambda^{d-6}\sim {\epsilon}^{7/3}\Lambda^{d-6}\sim D_4$
and disappears if $\tilde{D}$ vanishes.
Like our previous analysis, this can be interpreted as if the free energies
$h_1,..,h_4$ satisfy a linear equation and the noises have four-point
crosscorrelations given by
\begin{equation}
\langle f_1 f_2 f_3 f_4 \rangle \equiv D_4 \delta({\bf x}_1-{\bf x}_2)
\delta({\bf x}_1-{\bf x}_3)\delta({\bf x}_1-{\bf x}_4).
\end{equation}
We calculate $q_4$ from the relation
\begin{equation}
q_4=-{1\over t}{dF(D_4 t^{-\phi_4/z})\over D_4}|_{D_4=0}\sim t^{\Sigma_4}.
\end{equation}
We find $\phi_4=-(z-3d-4\chi+6-d)=4\epsilon$. Hence,
we obtain $\Sigma_4=(-4\epsilon-2)/2$, i.e., $q_4\sim t^{(-3\epsilon-2)/2}$.
\begin{figure}
\epsfxsize=6cm
\epsfysize=6cm
\centerline{\epsffile{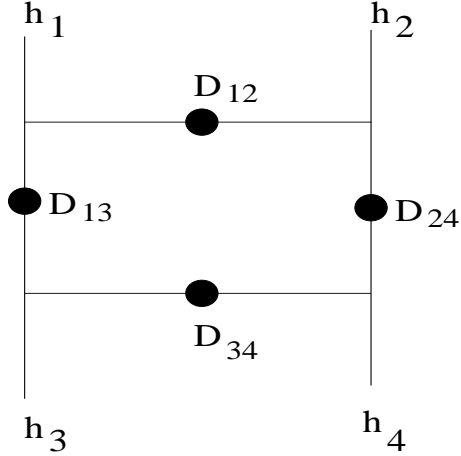}}
\caption{\small A schematic one-loop diagram contributing to 
$<f_1f_2f_3f_4>$. $D_{12}=D_{23}=D_{34}=D_{14}=\tilde{D}$.}
\end{figure}       

\subsection{effective interactions between  of $m$ polymers}
\begin{figure}
\epsfxsize=6cm
\epsfysize=6cm
\centerline{\epsffile{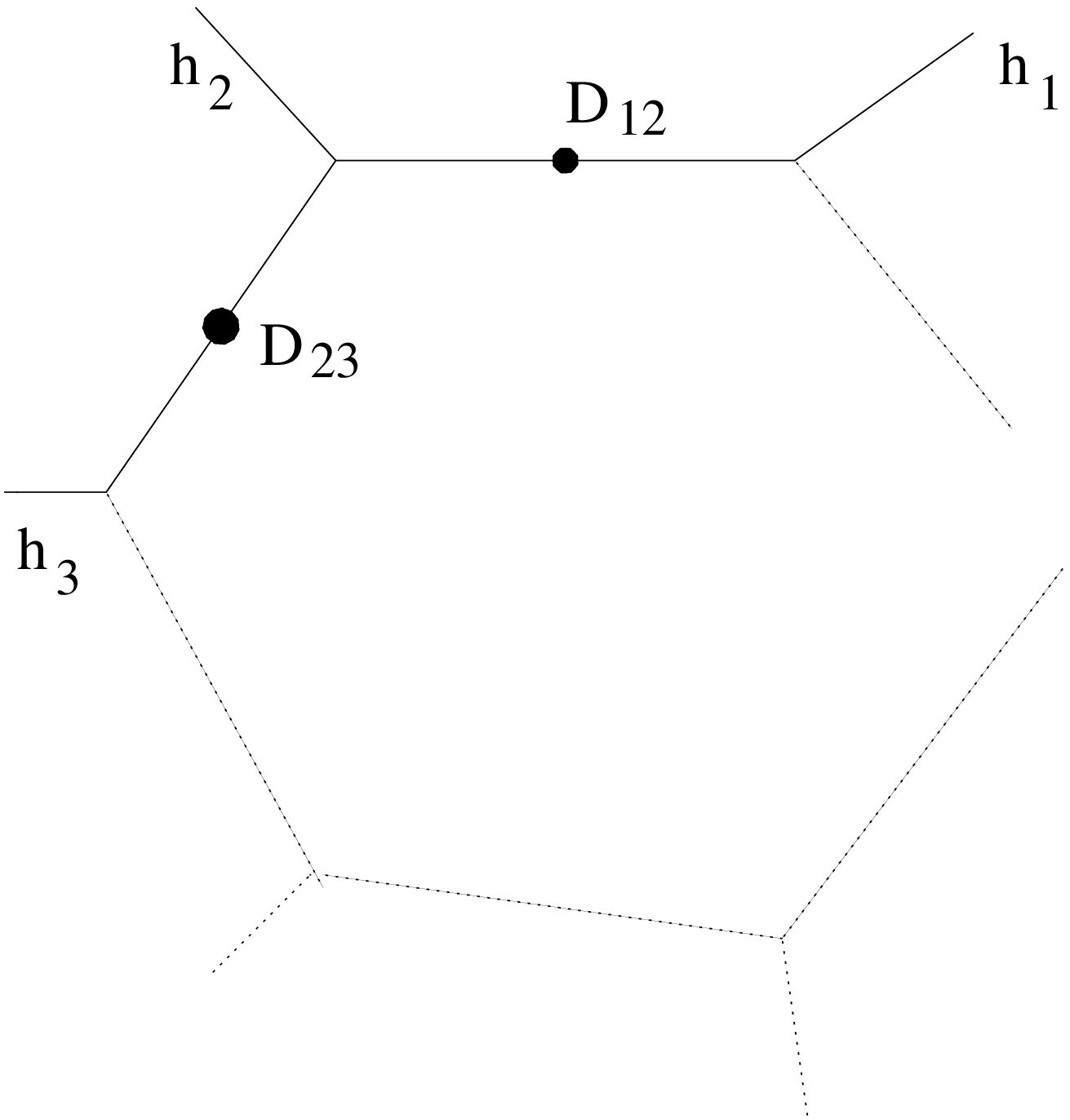}}
\caption{\small Diagramatic contribution to $<f_1f_2...f_m>$ upto one-loop
order.}
\label{mov}
\end{figure}

It is easy to convince oneself that any higher point correlation of the
random potential is non-zero. Notice that these $m$-body effective interactions
are built from the pairwise crosscorrelations. Since the two chain effective
interaction is attractive, these $m$-chain interactions constructed out of
that are also attractive.
Fig.(\ref{mov}) is a typical diagram which contributes at $O(\lambda^m)$
($\equiv$one-loop) to $\langle V_1...V_m\rangle$ or to 
$\langle f_1...f_{m}\rangle$. A nonzero value of this ensures phase transitions
involving $m$-polymers
The one-loop integrals scale as $\sim \Lambda^{\epsilon+4-2{m}}$.
We immediately obtain 
\begin{equation}
\phi_m=-[z-(m-1)d-m\chi+2m-4-\epsilon]=m\epsilon.
\end{equation}
This leads to
\begin{equation}
q_{m}\sim t^{\Sigma_{m}},\;\;\; \Sigma_{m}=[-(m)\epsilon-2]/2,
\label{sigab}
\end{equation}
at $d=2+\epsilon$.
Naturally the effective free energy contains all the interaction terms which
are generated due to disorder averaging. Equivalently the disorder mediates
interactions betweens arbitrary number of polymers.

AT $d=1$, along AX, $\phi_m=-1/2-m/2<0,\;(m>2)$, hence the effective $m$-point 
coupling disappears in the large length scale limit. Thus there is now 
$m$-body interaction in that limit. However, $\phi_2=0$,i.e., effective two
body attarctive interaction is marginal causing two polymers to attract.

\section{Transitions along the line YO}
So far we focussed on transitions at $g\sim \epsilon$, i.e., along the
line AX. But one could have followed a different path in the $(g-\tilde{g}$ plane: In
particular if one follows the line $g=0$ then the unstable fixed point $(0,\tilde{g}^*)$
gives rise to transitions with different exponents. Along this
line $g^*=0$, hence {\em individual} polymers are free (i.e., randomness of the medium
is irrelevant as far as their transverse fluctuations are concerned. These
fluctuations are still described by $z=2$ and $\chi=d/2$ in $d$-dimensions). 
However these free polymers still see attractive contact interactions
in presence of one another which causes these transitions. At dimension $d=2+\epsilon,\;
g^*=0,\;z=2$ and $\chi=d/2$; hence at this fixed point
\begin{eqnarray}
\phi_m&=&-[z-(m-1)d-m\chi+2m-4-\epsilon]=({3\over 2}m-1)\epsilon,\\
\Sigma_m&=&(\chi-\phi_m-z)/z=[{3\over 2}m\epsilon-2]/2.
\end{eqnarray}
Physically along this line polymers are individually free (described by 
Gaussian polymer exponents), no matter what the value of 
$\tilde{g}$ is. If the value of $\tilde{g}$ is higher than a critical value 
then Gaussian polymers attract each other due to contact interactions
induced by disorder between different them.

\section{Summary}
In this paper we have discussed how disorder generates attarcative
 interactions between two directed polymers in a disordered medium.
We show that due to these attarctive interactions the polymers undergo a 
binding-unbinding transition. We argue that in terms of the KPZ 
language this is due to
the crosscorrelations of noises in the two KPZ equations satisfied by the
respective free energies of the two polymers. We present a detailed analyis
of the phase diagram. The strength of the cross correlation (i.e., the strength
of the effective attaractive interaction) appears as a new parameter in the 
problem. This is relevant, in an RG sense, above a threshold value in any
space dimension $d$. In $d=2+\epsilon$ we get a new nontrivial unstable 
fixed point signalling a second order binding-unbinding transition. At
$d=2+\epsilon$ we get four different phases, two of which are new arising
due to the crosscorrelation only. Similarly in $1d$ two phases appear, one of 
them is new. These effects can be realised by
putting different kinds of polymers in the medium.  Notice that if the two
DPs are identical then $g=\tilde{g}$ and there is no independent variation of
$g$ or $\tilde{g}$. In that case the system is described by the line $g=
\tilde{g}$. It will be very interesting to examine this issue of disorder
induced phase transitions of DPs by using a variational replica approach
\cite{bouch} or a replica Bethe ansatz approach \cite{kar1}. In a typical 
replica calculation, the $n$-replica Hamiltonian $H_n$ becomes a function
of the replica-replica interaction terms. In the present case, in a replica
calculation $H_n$ will be the replica hamiltonian for $2n$ DPs. It will involve
`cross-replica' interaction terms arising due to the crosscorrelation of the
random potential. It will be interesting to see how the results obtained 
in this paper can be obtained in a replica approach.

\section{Acknowledgement}
The author wishes to thank S. M. Bhattacharjee for drawing his attention
to this problem and many discussions, and S. Ramaswamy,  
J. K. Bhattacharjee, and M. L\"{a}ssig for critical comments.

\end{document}